\input harvmac
\input epsf
\ifx\epsfbox\UnDeFiNeD\message{(NO epsf.tex, FIGURES WILL BE
IGNORED)}
\def\figin#1{\vskip2in}
\else\message{(FIGURES WILL BE INCLUDED)}\def\figin#1{#1}\fi
\def\ifig#1#2#3{\xdef#1{fig.~\the\figno}
\goodbreak\topinsert\figin{\centerline{#3}}%
\smallskip\centerline{\vbox{\baselineskip12pt
\advance\hsize by -1truein\noindent{\bf Fig.~\the\figno:} #2}}
\bigskip\endinsert\global\advance\figno by1}

\baselineskip 14pt

\def \s {\sigma}

\def \p {\phi}
\def \ha {\half}
\def \ov {\over}

\def \lr { \lref}
\def\ep{\varepsilon}

\def\dd {\partial }

\def\l{\lambda}

\def \k {\kappa}

\def\n{\noindent}

\gdef \jnl#1, #2, #3, 1#4#5#6{ { #1~}{ #2} (1#4#5#6) #3}

\def\np {  Nucl.  Phys. }
\def \pl { Phys. Lett. }

\def \prl { Phys. Rev. Lett. }
\def \pr  { Phys. Rev. }
\def \cqg { Class. Quant. Grav. }
\def \jmp { J. Math. Phys. }

\def \ijmp {Int. J. Mod. Phys. }

\lr \cghs {C. Callan, S. Giddings, J. Harvey and A. Strominger,
\pr D45 (1992) R1005. }
\lr \witt {E. Witten, \pr D44 (1991) 314. }
\lr \birdav {N.D. Birrell and P.C.W. Davies, {\it Quantum fields in
curved space} (Cambridge University Press, Cambridge, England, 1982). }
 \lr \chrisful {S.M. Christensen and S.A. Fulling, \pr D15 (1977) 2088. }
\lr \gidnel {S.B. Giddings and W.M. Nelson, \pr D46 (1992) 2486. }
\lr \hawk {S.W. Hawking, Commun. Math. Phys. 43 (1975) 199;
\pr D14 (1976) 2460.}
\lr \rustse{ J.G. Russo and A.A. Tseytlin, \np B382 (1992) 259.}

\lr \rst {J.G. Russo, L. Susskind and L. Thorlacius,
\pr D46 (1992) 3444; \pr D47 (1993) 533.  }

 \lr \hawk {S. Hawking, Commun. Math. Phys. 43 (1975) 199.  }
\lr \hooft { G. 't Hooft, Physica Scripta T36 (1991) 247;
 {\it Dimensional reduction in quantum gravity},
Utrecht preprint THU-93/26, gr-qc/9310006. }

\lr\thoo { G. 't Hooft,  \np B256 (1985) 727;    \np B335(1990) 138. }
\lr\hft{C. Stephens, G. 't Hooft and B.F. Whiting,
\cqg 11 (1994) 621. }

 \lr \membr {K. Thorne, R. Price and D. MacDonald, {\it Black holes:
the membrane paradigm} (Yale Univ. Press, New Haven, CT, 1986). }
 \lr \rstf  {J.G. Russo, L. Susskind and L. Thorlacius, \pl B292 (1992) 13.}
\lr \veil  {J.G. Russo, \pr D49 (1994) 5266.}
\lr\wald {R. M. Wald, {\it General Relativity} (University of
Chicago Press, Chicago, 1984).}
 
\lr \rst {J.G. Russo, L. Susskind and L. Thorlacius,
\pr D46 (1992) 3444; \pr D47 (1993) 533.  }
\lr\verli{H. Verlinde and E. Verlinde, \np B406 (1993) 43;
K. Schoutens, H. Verlinde and E. Verlinde, \pr D48 (1993) 2670.
}
\lr\thoof { G. 't Hooft,  \np B335(1990) 138.  }

\lr\jacob {T. Jacobson, \pr D44 (1991) 1731.}
\lr\susskin {  L. Susskind, L. Thorlacius and J. Uglum,
\pr D48 (1993) 3743.}

\lr\ssk{ L. Susskind, \prl 71 (1993) 2367. } 

\lr\sussk {L. Susskind,  \jmp 36 (1995) 6377. }
  
\lr\boulw {D.G. Boulware, \pr D11 (1975) 1404. }
\lr\vercom{ 
Y. Kiem, H. Verlinde and E. Verlinde,  \pr D52 (1995) 7053.}
\lr \beken {J.D. Bekenstein, \pr D7 (1973) 2333; D9 (1974) 3292.  }
\lr\giddings{S. Giddings, \pr D46 (1992) 1347.}
\lr\hori{J.G. Russo, \pl B359 (1995) 69.}
\lr\black{J.G. Russo, {\it Model of black hole evolution}, hep-th/9602124.}
\lr\amati{ D. Amati, M. Ciafaloni and G. Veneziano,  \ijmp A3 (1988) 1615;
\np B403 (1993) 707.}

\baselineskip8pt
\Title{\vbox
{\baselineskip 6pt{\hbox{CERN-TH/96-34}}{\hbox
{SISSA/27/96/EP}}{\hbox{hep-th/9602125}} {\hbox{
   }}} }
{\vbox{\centerline {   Black holes   by analytic continuation
     }
}}

\vskip -20 true pt

\centerline  { D. Amati   }

 \smallskip \smallskip

\centerline{\it  SISSA, Trieste, Italy }
\smallskip

\centerline{\it    and INFN, Sezione di Trieste }

\bigskip

\centerline {and}

\bigskip

\centerline{  J.G.  Russo  }

\smallskip\smallskip
\centerline{\it  Theory Division, CERN}
\smallskip

\centerline{\it  CH-1211  Geneva 23, Switzerland}

\bigskip\bigskip\bigskip
\centerline {\bf Abstract}
\medskip
\baselineskip8pt
\noindent

In the context of a two-dimensional exactly solvable model,
  the dynamics of quantum black holes
is obtained by analytically continuing the description of the regime
 where no black hole is formed. The resulting 
spectrum of outgoing radiation departs   from the
one predicted by the Hawking model in the region where the  outgoing modes 
arise from the horizon with Planck-order frequencies.
This occurs early in the evaporation process,
and the resulting physical picture is unconventional.
The theory predicts that 
black holes will only radiate out  an   energy of Planck mass order, 
stabilizing after   a transitory period.   
The continuation from a regime without black hole formation --accessible in
the 1+1 gravity theory considered-- is implicit in an $S$-matrix approach
and suggests in this way a possible solution to the problem of 
information loss.

\Date {February 1996}

\noblackbox
\baselineskip 16pt plus 2pt minus 2pt


\vfill\eject

\newsec{ Introduction}

It has   often been advocated \refs {\thoo , \amati } that the study of scattering of matter and radiation in a quantum gravity theory should solve the conflict between classical black-hole solutions and quantum mechanics, which leads to 
information loss \hawk . 
The mere existence of an $S$-matrix below the threshold of 
black hole formation would be enough to exhibit, through its analytic
structure, eventual thresholds for the creation of new objects, and to describe,
through analytic continuation, the physics above them in a unitary 
framework.

 By studying a semiclassical   solvable model in which the black hole
evolution can be explicitly investigated,
we will see that   analytic continuation (from below
the threshold of black hole formation to above it) 
completely determines the structure of the theory in the
regime in which black holes are formed.
The model   is the two-dimensional dilaton gravity with matter, RST
  \rst  , which  represents a toy model for   spherically symmetric
infalling shells in four-dimensional gravity.
Due to quantum effects there is a threshold on the incident matter 
energy density under which there is no  black hole formation.

We shall adopt the usual boundary conditions below the threshold, 
so the subcritical regime will be   as in  the  RST model.
It will then be shown that the corresponding  outgoing energy-momentum tensor  can be straightforwardly continued 
above the critical incoming energy-density flux.
The semiclassical supercritical treatment that would give rise to the same outgoing radiation requires a boundary  at the apparent horizon
(this is at variance with the standard boundary on the singularity). As a result, a very unconventional picture appears. In particular,
Hawking radiation stops early in the evaporation process
and a stable macroscopic black hole remains in the final state,
thus avoiding the information loss.
This goes in the direction advocated by Giddings \giddings\ as a possible
solution of the information loss problem.

   In sections 2 and 3 we briefly review the model of \rst\ and the 
subcritical regime. 
The obtained outgoing energy-momentum tensor may be continued beyond 
the threshold
--section 4--
and we discuss in section 5
which      boundary conditions   would reproduce it.
Section 6 generalizes the preceding results to other infalling distributions
 of interest.
 In  Section 7 we summarize the  physical picture
and discuss the origin
of the differences with preceding
treatments.

\newsec{ Semiclassical dilaton gravity}
 The semiclassical action of the RST model  (which includes the one-loop quantum anomaly) is given by
\eqn\uno{
S ={1\ov 2\pi} \int d^2 x \sqrt{-g}\bigg[ e^{-2\phi}\big( R+
4(\nabla \phi)^2  +4\l^2  \big) 
- \ha \sum_{i=1}^N  (\nabla f_i)^2  
 } 
$$
- \k \big(  2  \phi R +  R (\nabla^2)^{-1} R \big)\bigg] \ ,
\ \ \ \ \k={N\ov 48}\ .
$$
In the conformal gauge  $g_{\pm\pm}=0$, $g_{+-}=-{1\ov 2}
e^{2\rho}$ , the action is simplified by introducing new fields
$\chi $, $\Omega $, related to $\rho $ and $\phi $ by
\eqn\due{
\chi = 4\k\rho + e^{-2 \phi}-2 \k\phi  \ ,\ \ \ 
 \Omega=e^{-2\phi}+ 2\k \phi\ \     .
}
Then action \uno\ takes the   form
\eqn\free{
S={1\ov{\pi}}\int d^2 x \big[ {1\ov 4\k }
(-\dd_+\chi \dd_-\chi +\dd_+\Omega\dd_-\Omega) +
\l^2 e^{{1\ov 2\k }(\chi - \Omega)} + {1\ov 2}\sum_{i=0}^{N}
\dd_+ f_i \dd_- f_i \big] \ ,
}
with the constraints (corresponding to the $g_{\pm\pm }$ equations of motion
of action \uno )
\eqn\quatt{
 t_{\pm}(x^\pm )={1\ov 4 \k}(-\dd_\pm \chi\dd_{\pm}\chi 
 + \dd_{\pm}\Omega\dd_{\pm}\Omega ) +  \dd_{\pm}^2 \chi +
{1\ov 2} \sum_{i=0}^{N} 
\dd_{\pm}f_i \dd_{\pm}f_i \ .}
The functions $t_\pm (x^\pm )$  are determined by boundary conditions.

Let us consider a general distribution of incoming matter: 
$$
T_{++}(x^+)={1\ov 2}\sum_{i=0}^{N}
\dd_+ f_i \dd_+ f_i \ .
$$
 In the Kruskal gauge, $\chi=\Omega $, the solution  to the semiclassical equations of motion and the constraints  is given  by
\eqn\sette{\Omega=\chi=-\l^2 x^+ 
\big( x^- + \l^{-2} P_+(x^+) \big) -{\k}\ln (-\l^2 x^+x^- )
+\l^{-1} M(x^+)   \ ,
}
where $M(x^+)$ and $P_+(x^+)$ physically represent the total energy and total Kruskal momentum
of the incoming matter at advanced time $x^+$ :  
$$
P_+(x^+)=\int_0^{x^+}dx^+T_{++}(x^+) \ ,\ \ \ 
M(x^+)=\lambda \int_0^{x^+} dx^+ x^+ T_{++}(x^+) \ .
$$
In the particular case $T_{++}=0$, eq. \sette\ reduces to the familiar linear dilaton vacuum,
\eqn\ldv{
e^{-2\phi }=e^{-2\rho }= -\l ^2 x^+x^- \ .
}
In Minkowski coordinates $\s^\pm $, 
$\l x^\pm =\pm e^{\pm \l\s^\pm } $, one has $ds^2=-d\tau^2+d\s^2$, 
$\phi=-\l\s  ,\ \s^\pm=\tau \pm \s $.

The curvature scalar of the   geometry, 
$R=8 e^{-2\rho } \dd_+\dd_- \rho $, can be conveniently written as
 \eqn\seuno{
R = 8e^{-2\rho} {1\ov  \Omega ' (\phi)}
( \dd_+\dd_- \chi - 4\dd_+\phi\dd_-\phi e^{-2 \phi}) \ .
}
In this form we see that, generically, there will be a curvature singularity at 
 $\phi=\phi_{\rm cr}=-{1\ov 2}\ln {\k}$ where $\Omega '(\phi )=0$.

As observed in ref. \rst , there are two different regimes, according to whether
$T_{++}(x^+)$ is smaller or greater than a critical flux
\eqn\cflux{
T_{++}^{\rm cr}(x^+)= {{\k}\ov{x^{+2}}} \ .
}
Note that the existence of the threshold is a quantum effect. Indeed,
$\k $ is proportional to $\hbar $ (here we  have set $\hbar =1$)  
and thus $T_{++}^{\rm cr}$ vanishes as $\hbar \to 0$.
Using eq. \sette\ it can be seen that the line 
$\Omega=\Omega_{\rm cr}$ ($\equiv \Omega (\phi_{\rm cr} )$ )
  is time-like 
 if $T_{++}(x^+)< T_{++}^{\rm cr}(x^+) $, and it becomes space-like 
  as soon as $T_{++}(x^+) > T_{++}^{\rm cr}(x^+) $.

\newsec{ Subcritical regime}
 
In order to investigate the analytic continuation of the subcritical
regime to  a supercritical regime, it is convenient to
explore in more detail the  subcritical theory of ref.~\rst .
Let us assume that  the geometry is originally the linear dilaton
vacuum, and there is an   incoming energy density flux 
$T_{++}(x^+)< T_{++}^{\rm cr}(x^+) $,
which is different from zero for $x^+_0<x^+<x^+_1$.
  Let us define region (i) as  $x^-<x^-_0$, $ x^-_0=-\k/(\l^2 x^+_0)$, 
and region (ii) 
as that between $x_0^-$ and $x^-_1$ (see fig. 1).

\ifig\fone{ 
Kruskal diagram in the subcritical regime.
}{\epsfxsize=5.0cm \epsfysize=5.5cm \epsfbox{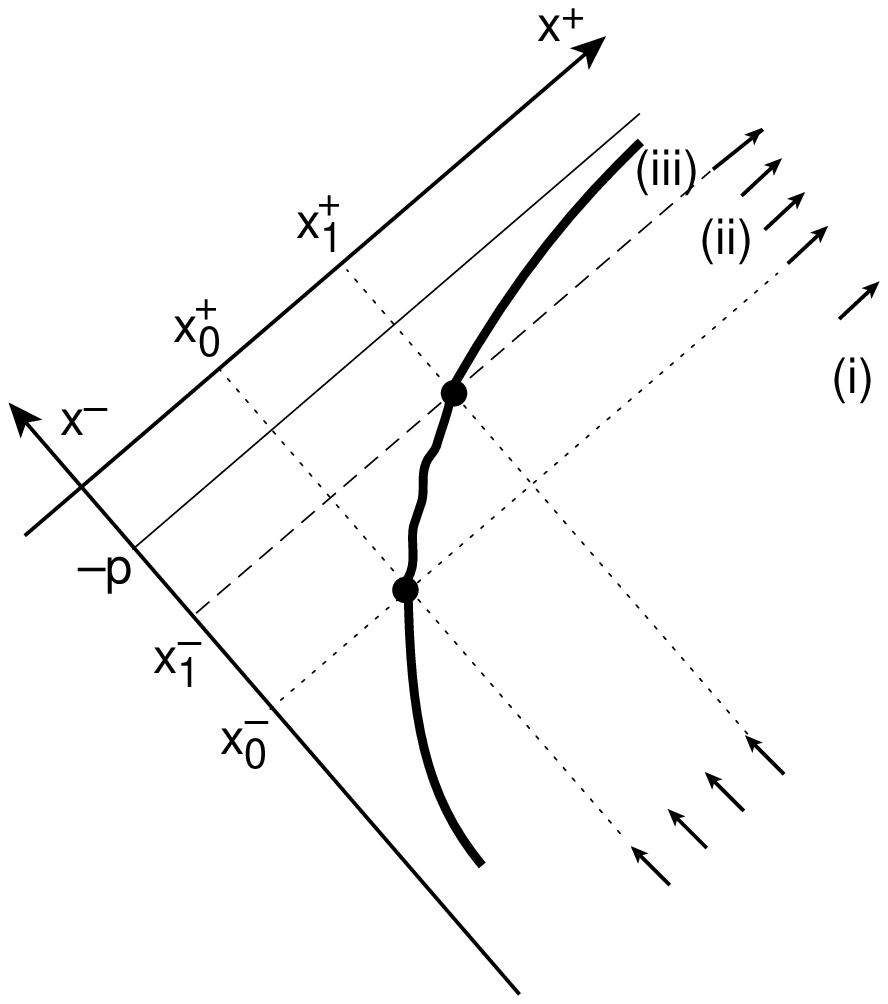}}

In region (i), the solution is given  by eq. \sette , which is completely specified by the asymptotic
boundary conditions and by demanding a continuous matching with the linear dilaton vacuum in the infalling line.
In region (ii) the boundary $\phi =\phi _{\rm cr }$ is time-like and  
boundary conditions are needed in order to determine the evolution.
Continuity along the line $x^-=x^-_0$ requires that the solution in region
(ii) be of the form
\eqn\ssss{
\Omega^{\rm (ii)}(x^+,x^-)=\Omega^{\rm (i)}(x^+,x^-)+F(x^-)\ ,
\ \ \ \ F(x^-_0)=0\ .
}
 The ``reflecting"   RST boundary conditions  
 follow from the requirement of finite curvature on the boundary line. Indeed, from eq. \seuno , we see that  in order to have finite curvature at  the line $\Omega '(\phi )=0$ it is necessary that 
\eqn\neces{
\dd_+\p \dd_-\p \bigg|_{\phi=\phi_{\rm cr}} =-{\l ^2\ov 4 \k}\ \ ,
}
where we have used the equation of motion (in the gauge $\chi=\Omega $)  $\dd_+\dd_-\chi=- \l ^2 $ . Equation \neces\ implies, in particular
(see eq. \due ),
\eqn\oooo{
\dd_+\Omega \bigg|_{\phi=\phi_{\rm cr}}=
\dd_-\Omega \bigg|_{\phi=\phi_{\rm cr}}=0 \ .
}
As a result, the function $F(x^- )$ is determined to be
\eqn\ffff{
F(x^-)={\k} \ln \bigg[ {x^- \ov x^-+\l^{-2} P_+(\hat x^+) } \bigg]
- \l^{-1} M(\hat x^+)\ , }
where $\hat x^+=\hat x^+(x^-)$ is the boundary curve given by
\eqn\bound{
-\l^2 \hat x^+ 
\big( x^- + \l^{-2} P_+(\hat x^+) \big) = {\k}\ .
}
{}Finally, in region (iii), the geometry is matched with the vacuum:\
\eqn\ommmm{
\Omega ^{\rm (iii)}=\chi ^{\rm (iii)}=-\l ^2 x^+(x^-+ p) -
{\k }\ln \big[ -\l ^2 x^+ (x^-+p ) \big]\ ,\ \ \ p\equiv \l^{-2}P_+(x^+_1)\ .
}
In Minkowski  coordinates, 
$\l (x^-+p) =- e^{-\l\s^- } ,\ \l x^+=e^{\l\s^+ }$, this simply becomes $ds^2=-d\s^+ d\s^- $,   $\phi=-\l\s $. 

 The outgoing   energy density fluxes
measured by an out
observer can be found from the constraints. They are given by 
\eqn\ttti{
T^{\rm (i )}_{--}(x^- )= {\k } \bigg[ {1\ov (x^-+p)^2 } -{1\ov {x^-}^2}\bigg]\ ,
}
\eqn\tttii {
T^{\rm (ii )}_{--}(x^- )= {\k }   {1\ov (x^-+p)^2 } - {\l^4  \ov 
{\k\ov \hat x^{+2}} -   T_{++}(\hat x^+ ) }\ ,\ \ \
}
\eqn\ttttt{
 \ T^{\rm (iii)}_{--}=0\ .
}
  The radiation energy emitted between times $\s^-_1$
and $\s^-_2$ is given by the integral 
 $$
E= \int _{\s^-_1 }^{\s^-_2}  d\s^- T_{\s^-\s^-}=-\l 
\int  _{x^-_1 }^{x^-_2}dx^- (x^-+p)\ T_{--}\ .
$$
Thus the total radiated energies in regions (i) and (ii) are
\eqn\energi{
E^{\rm (i)}_{\rm out}= 
-\l \int _{-\infty}^{x_0^-} dx^- (x^-+p)\ T_{--}^{\rm (i)}
= -{\l \k p\ov x^-_0}- {\l \k }\ln \big( 1 + { p\ov x^-_0 } \big)\ ,
}
\eqn\energii{
E^{\rm (ii)}_{\rm out}= 
-\l \int _{x_0^-}^{x_1^-} dx^- (x^-+p)\ T_{--}^{\rm (ii)}
=m+   {\l \k p\ov x^-_0} +  {\l \k }\ln \big( 1+ { p\ov x_0^-} \big)\ ,
}
where $m\equiv M(x_1^+)$ is the total ADM energy of the initial configuration.
The coordinate $x_0^-$ is related to the time $x^+_0$, at which the incoming flux begins,
by $x^-_0=-{\k \ov \l ^2 x^+_0 }$.
For $p\ll |x_0^-|$ (``low-energy" fluxes), one 
has $E^{\rm (i)}_{\rm out}\ll m$, 
$E^{\rm (ii)}_{\rm out}\sim m $, that is most of the energy comes out by pure
reflection on the space-time boundary. For $p\cong -x^-_0$, $p< |x_0^-|$, the logarithm becomes large and negative so that the   energy radiated 
in region (ii) is negative.

Note that it is possible to have $T_{++}>T^{\rm cr}_{++}={\k\ov {x^+}^2}$
and yet $p<-x^-_0$. This means that the threshold given by the
singularity of the logarithm in eqs. \energi , \energii\ is not in general the threshold for black hole formation.
To see this explicitly, let us consider the simplest case in which the incoming
energy-density flux is constant  in Minkowski coordinates,
so that in Kruskal coordinates it reads
\eqn\eecc{
T_{++}(x^+)={\ep \ov {x^+}^2}\ .
}
Whence 
\eqn\ppmm{
\l ^2 p= P_+(x^+_1)= \ep \bigg( {1\ov x_0^+} - {1\ov x_1^+} \bigg)
\ ,\ \ \ \ m=\l \ep  \ln ( x^+_1/x^+_0)\ .
}
 {}For $x^+_1$ close enough to $x^+_0$ (more precisely, for
$x_1^+<x^+_0/(1-{\k\ov \ep })\ ,\ \ep >{\k }$)
we can have $1+{p\ov x^-_0} >0$  even above the threshold for black hole
formation, i.e. with $\ep >{\k }$.
 
\newsec{ The intermediate regime }

As mentioned before,
eqs. \energi\ and \energii\  can be continued above the threshold
without encountering
any singularity  up to $p= |x^-_0|$, $p\equiv \l^{-2} P_+(\infty )$, where the logarithmic singularity appears.
We shall call this the intermediate regime, i.e. the case when  
$p< |x^-_0|$ and $T_{++}>T^{\rm cr}_{++}$ for some $x^+$,
as opposed to  the ``supercritical" regime where $p> |x^-_0|$.
The former describes small ``Planck-size" black holes, whereas
the latter includes macroscopic black holes
(the classical  picture is approached for $p\gg |x^-_0| $~).

\ifig\fdos{ 
Intermediate regime.
}{\epsfxsize=5.0cm \epsfysize=5.2cm \epsfbox{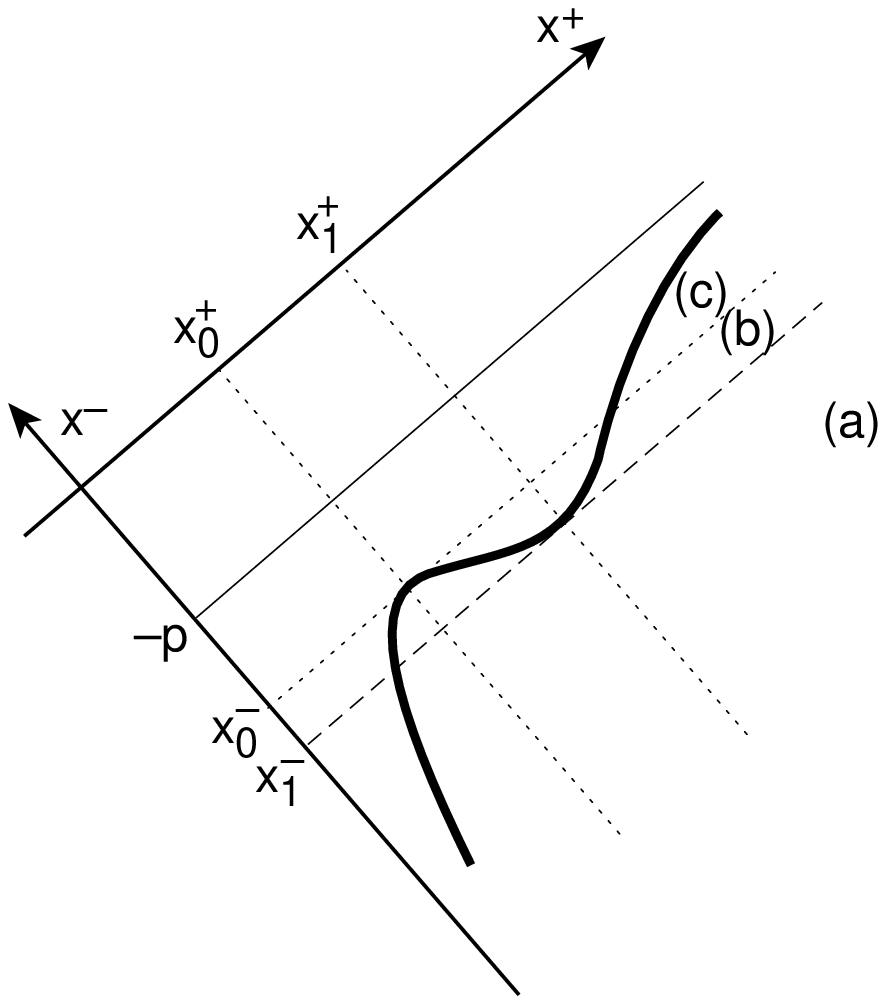}}

 The geometry is
exhibited in fig. 2 for the case the incoming energy density is larger
than the critical one in the whole range $x^+_0<x^+<x^+_1$.
Region (a) is defined as $x^-<x^-_1$,   region (b) as $ x^-_1<x^-<x^-_0$,
and region (c)   as $x^->x^-_0$.
Figure 2 can be understood as a deformation of fig. 1.
In this process  region (ii) and part of (i) of fig. 1 are superposed into
region (b) of fig. 2. Region (c) is part of region (iii), so that
$T^{\rm (c)}_{--}=T^{\rm (iii)}_{--}=0 $.  
It is thus convenient to split the  integral \energi\  as
\eqn\enei{
E^{\rm (i)}_{\rm out}=  -{\l \k p\ov x^-_0}- \l \k \ln \big( 1 + { p\ov x^-_0 } \big)
=E^{\rm (a)}_{\rm out}+E^{\rm (ib)}_{\rm out}\ ,
}
where
\eqn\enea{
E^{\rm (a)}_{\rm out}=
-\l  \int _{-\infty}^{x_1^-}   dx^- (x^-+p) \ T_{--}^{\rm (i)}
= -{\l \k p\ov x^-_1}- {\l \k }\ln \big( 1 + { p\ov x^-_1} \big)\ ,
}
\eqn\eneub{
E^{\rm (ib)}_{\rm out}=-\l   \int _{x_1^-}^{x_0^-}  dx^- (x^-+p) \ T_{--}^{\rm (i)}= -{\l \k p\ov x^-_0}+{\l \k p\ov x^-_1} - {\l \k }\ln
{ \big( 1 + { p\ov x^-_0 } \big)\ov  \big( 1 + { p\ov x^-_1 } \big)}\ .
}
 The first integral gives the energy radiated in region (a) of fig. 2.
The second integral   contributes to the radiation in region (b). The total energy
in region (b) is obtained by adding $E^{\rm (ii)}_{\rm out}$.  Since 
now $x_1^-< x^-_0$, it is convenient to write  this integral in
the following way:
 \eqn\eneii{
E^{\rm (ii)}_{\rm out}= 
-\l \int _{x_1^-}^{x_0^-} dx^- (x^-+p)\ \big( - T_{--}^{\rm (ii)}\big)
=m+   {\l \k p\ov x^-_0} +  {\k\l }\ln \big( 1+ { p\ov x_0^-} \big)\ ,
}
so that
 \eqn\eneb{
 E^{\rm (b)}_{\rm out}= E^{\rm (ib)}_{\rm out}+
E^{\rm (ii)}_{\rm out}  
=-\l \int _{x_1^-}^{x_0^-} dx^- (x^-+p)\   \tilde T_{--}^{\rm (b)}\ ,
\ \ \ \ \ \tilde T_{--}^{\rm (b)}\equiv T_{--}^{\rm (i)}-T_{--}^{\rm (ii)}\ ,
}
\eqn\enebb{
 E^{\rm (b)}_{\rm out}=
m + {\l \k p\ov x^-_1} + {\k\l }\ln \big( 1+ { p\ov x_1^-} \big)\ .
}
Clearly,  $ E^{\rm (a)}_{\rm out}+ E^{\rm (b)}_{\rm out}=m$, so that the whole incoming energy has been radiated (see eqs.~\energi , \energii ).
This means that these black holes evaporate completely.

It should be noticed that in the region (b)
(i.e. in    the region    in causal contact with the apparent horizon)
 the    $T_{--}$  arising in the RST formalism
  does not coincide with the  straightforward
continuation of the subcritical formulas given by eq. \eneb .
Indeed, in RST the energy-momentum tensor keeps being 
$T_{--}^{\rm (i)}$ until the geometry is matched with the vacuum.
Although in both cases the original energy is completely radiated, 
the structure of the outgoing energy-density flux in the two models is  different.


\newsec{Apparent horizon as a boundary}

In the subcritical regime the boundary conditions \oooo\ 
  or, equivalently,
 \eqn\bcca{
\Omega = \Omega_{\rm cr }\ ,\ 
}
\eqn\bccb{
\dd_+\Omega=0\ ,
}
can be implemented simultaneously
on some line.  Above the threshold the line defined by eq. \bcca\ is necessarily
different from the line defined by eq. \bccb . 
The  usual choice  is to define the boundary line by $\Omega=\Omega _{\rm cr}$, since it is
on this line that the curvature is singular. 
This  leads to the black hole evolution   described in \rst \ which,
although it
reproduces the standard Hawking model of gravitational collapse,   does
   not correspond to the
analytic continuation of the subcritical regime. We will now
show that the other option, namely imposing boundary conditions
at $\dd_+\Omega (x^+,x^-)=0$ (the apparent horizon), 
will reproduce the results that were previously obtained   by
a simple continuation of the subcritical formulas.

 Let us assume that the incoming supercritical energy-density flux $T_{++}(x^+)$
starts at $x_0^+$, and it is turned off at a later time $x^+_1$
(a more general situation is discussed in sect. 6).
In region (a) the geometry will be given by eq. \sette .
The boundary $\dd_+\Omega =0$ becomes time-like for $x^->x_1^-$, and boundary conditions
are needed in order to determine the evolution of the geometry in region (b)
(see figs. 2, 3). Continuity along the line $x_1^-$ requires that
\eqn\ssss{
\Omega^{\rm (b)}(x^+,x^-)=\Omega^{\rm (a)}(x^+,x^-)+F(x^-)\ ,
}
with $F(x^-_1)=0$. We need to generalize the expression \ffff\ for the
case when there is   some   energy stored in the geometry  by the time the
boundary becomes time-like. The form of $F(x^-)$ in the subcritical regime
suggests the choice (the general structure will be  clear in sect. 6)
\eqn\gggg{
F(x^-)={\k } \ln \bigg( {x^- +\l^{-2}P_+( u) \ov 
x^- + \l^{-2}P_+(x_1^+) } \bigg)
+ \l^{-1} M(u)- \l^{-1} M(x^+_1)\ ,
}
with $u(x^-)$ given by the 
branch $x^+_0<u<x_1^+$ of the solution to the equation
 \eqn\uuuu{
-\l^2 u
\big( x^- + \l^{-2} P_+(u ) \big) = \k  \ .
}
We will find   that given the boundary condition \ssss\ with \gggg, \uuuu,
then  eqs. \enea, \eneb\ are reproduced (in particular, this means that
this boundary condition conserves energy). The formulas for the outgoing 
fluxes will be identical to those obtained by direct extrapolation from
the subcritical regime.

\ifig\ftre{      Apparent horizon in the
supercritical regime.    
}{\epsfxsize=5.0cm \epsfysize=5.0cm \epsfbox{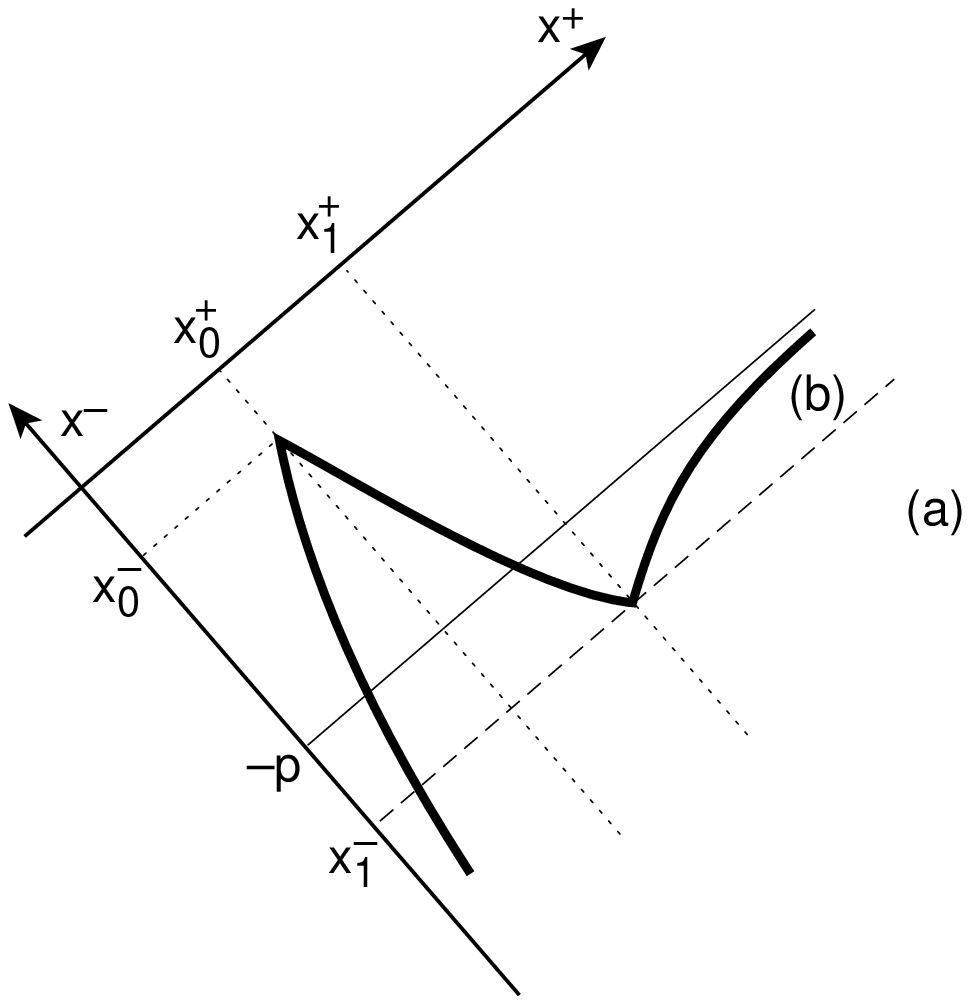}}

Let us note that the matching between regions (a) and (b) is smoother than in   the case of  \rst , i.e. there is no outgoing shock wave:
\eqn\pop{
T_{--}(x^-_1+\epsilon )-T_{--}(x^-_1-\epsilon )
= - {dF\ov dx^-} \delta (x-x^-_1)=0\ ,
}
since (see eqs. \gggg, \uuuu )
$$
{dF\ov dx^-}={\k } \bigg[ {1\ov x^-+\l^{-2}P_+(u)} -{1\ov x^-+p }\bigg]
$$
vanishes at $x^-=x^-_1$.

\subsec{Intermediate regime}

 Let us first consider the intermediate regime. In region (c), defined
by $-p>x^->x^-_0$, the geometry is matched with the linear dilaton vacuum, i.e.
\eqn\ldviii{
\Omega^{\rm (c)}(x^+,x^-)= -\l ^2 x^+(x^-+p) -{\k } \ln \big( -\l ^2 
x^+(x^-+p) \big)\ .
}
The energy-momentum tensor in the different regions are  found to be  
($T_{--}=-\dd_- ^2\Omega +t_-(x^-) ,\ t_-(x^-)=\k/(x^-+p)^2 $)
\eqn\ttta{
T^{\rm (a )}_{--}(x^- )= {\k } \bigg[ {1\ov (x^-+p)^2 } -{1\ov {x^-}^2}\bigg]
\ ,
}
\eqn\tttb{
T^{\rm (b )}_{--}(x^- )= \l ^2 {du\ov dx^-} - {\k \ov {x^-}^2 }\ ,\ \ \ \ \ 
T^{\rm (c)}_{--}(x^- )=0\ .
}
In particular, we note that, since $ u'=\l ^2 ( {\k\ov u^2}-T_{++}(u) )^{-1}<0$
 (the flux is above the critical flux), the outgoing flux in region (b)
carries negative energy.

Since in region (a) the solution was not modified, one has 
$T^{\rm (a )}_{--}=T^{\rm (i )}_{--} $ (see eq. \ttti ). Now we note the surprising relation (see eqs. \ttti , \tttii , \eneb )
\eqn\eeee{ 
T^{\rm (b )}_{--}=T^{\rm (i )}_{--}-T^{\rm (ii )}_{--}\equiv \tilde 
T^{\rm (b )}_{--}\ .
}
Thus the outgoing energy-momentum tensor in this theory with $\dd_+\Omega =0$ as a boundary coincides with the extrapolation of the subcritical
energy momentum tensor beyond the threshold for black hole formation,
  indicating that it is the theory
defined with the boundary at the apparent horizon  that
represents the analytic
continuation of the subcritical regime.

\subsec{Supercritical regime}

Let us now proceed by considering the case $ p>|x^-_0|$ (fig. 3).\foot{
Note that possible discontinuities in $T_{++}(x^+) $ produce  
discontinuities  in the derivative of the curve (such discontinuities 
can of course be present in all regimes).}  
The energy-momentum tensor in region (b)  can either be obtained
by analytic continuation or by using eqs. \ssss, \gggg, and it
will   be  given by eq. \tttb ,
just as in the intermediate regime.
The final $\tau \to \infty $ 
geometry for a time-like observer is obtained by taking the limit
$x^+\to \infty $ and $x^-\to -p $ in   eqs. \ssss , \gggg\ 
(recall $2\l \tau = \ln x^+/|x^-+p|$~)
\eqn\lmviii{
\Omega^{\rm (b)}(x^+,x^-) \bigg|_{\tau\to \infty }= -\l ^2 x^+(x^-+p) - \k  \ln \big( -\l ^2 
x^+(x^-+p)\big) +{m_{ f} \ov \l } \ ,
}
\eqn\masa{
m_{ f}=  M(x^+_2) + {\l\k } \ln \big( 1- {P_+(x^+_2)\ov \l ^2 p} \big)\ ,
\ \ \ \ 
x_2^+\equiv u(-p)\ .
}
This  is   a static geometry   with ADM mass equal 
to $m_f$. In the whole of region (b), where 
$-\l ^2 x^+(x^-+p)>\k$, the logarithmic term can be  neglected and the
geometry is essentially the same as   the classical black hole geometry.
The logarithmic term is only significant close to the   line $x^-=-p$, where
there is a singularity. However, this is beyond  the boundary at the apparent horizon.

Let us check that energy is conserved. We now obtain by explicit integration:
\eqn\eeee{
  E^{\rm (a)}_{\rm out}=-\l \int_{\-\infty }^{x^-_1} dx^- (x^-+p) \ T_{--}^{\rm (a)} =  -{\l \k p\ov x^-_1}- {\l \k }\ln \big( 1 + { p\ov x^-_1} \big) \ ,
}
\eqn\eennn{
 E^{\rm (b)}_{\rm out}=-\l \int _{x_1^-}^{-p} dx^- (x^-+p) \ T_{--}^{\rm (b)} =  m  -M(x^+_2) + {\l \k p\ov x^-_1} + {\l\k } \ln {\big(
1+{p\ov x_1^-} \big) \ov \big( 1- {P_+(x^+_2)\ov \l ^2 p} \big) }\ ,
}
so that 
\eqn\conserva{
 E^{\rm (a)}_{\rm out}+ E^{\rm (b)}_{\rm out}=
m- M(x^+_2) +{\l\k } \ln \big(  {\l^2 x^+_2p \ov \k } \big)
=m-m_{ f}  \ ,
}
where we have used the relation 
$\l ^2 x^+_2\big( p-\l^{-2} P_+(x^+_2)\big) =\k $.
Thus energy is indeed conserved, and the total radiated energy is positive definite, since 
$$
m- M(x^+_2) = \lambda \int_{x^+_2} ^{x^+_1} dx^+ x^+ T_{++} 
 >0\ ,
$$
and $\ln \big(  {\l^2 x^+_2p \ov \k } \big)>0 $.
Indeed,
$\l^2 x^+_2p / \k > \l^2 x^+_0 p / \k= p/|x^-_0|$, with
$p/|x^-_0|>1$ in the supercritical regime.

Let us estimate the mass $m_f$ of the remaining black hole.
For a ``macroscopic" black hole, i.e. with $p \gg |x^-_0|$,
 it is clear that $M(x^+_2), P_+(x^+_2)$ will not differ much from 
$M(x^+_1), p\equiv P_+(x^+_1)$, since $x^+_2\cong x^+_1$ (see fig. 3).
We can therefore anticipate that $m_f \cong M(x^+_1)\equiv m$.
This means that very little energy has been radiated
and the  final black hole will have a mass similar to the total
imploding energy. 
This is very different from the standard picture of Hawking evaporation. 
To be   explicit, let us consider two extreme cases, namely the case
of a constant energy-density flux falling in for a long time,
and the case of a shock-wave collapse.
Using eqs. \eecc , \ppmm \ we find for the former  
$x_2^+=x^+_1 \big( 1- {\k\ov \ep } \big)$, and
\eqn\etot{
 E^{\rm (a)}_{\rm out}+ E^{\rm (b)}_{\rm out}={\l\k } \ln 
{\ep\ov \k}
\big({x^+_1\ov x^+_0 } -1 \big)  +\l (\ep -\k  ) \ln {\ep \ov \ep- \k  }\ .
}
For ${x^+_1\ov x^+_0 }\gg 1$ we get 
$$
m-m_f={\k\ov \ep } m\ .
$$
Since $p\gg |x^-_0|$ implies $\ep \gg \k $,  this is a small quantity.
The total radiated energy in the opposite limit of a shock wave 
can be found by using eq. \conserva\ and the fact
that, for a shock wave, $p=m/\l^3x_0^+ $. This gives
\eqn\etot{
m_f=m-\l\k \ln {m\ov \l\k } \ .
}
While the radiated energy   logarithmically increases  with $m$, 
the ratio $m_f/m \to 1$ as $m\to \infty $.

\subsec{Outgoing energies}

The energies radiated in region (i) of fig. 1 and in region (a) of figs. 2
and  3
are positive definite, since they are the integral of a positive-definite quantity
(see eq. \ttti ). We have also seen in the previous subsection that the total radiated energy is positive definite.
In the subcritical regime --as mentioned in sect. 3-- the energy in region
(ii) can be positive or negative, depending on the
characteristics of the incoming flux. This will be clear from the 
examples that we give below.  
As pointed out after eq. \tttb , the energy $E^{\rm (b)}_{\rm out}$
 is negative definite, being the integral of
a negative-definite quantity (see eq. \tttb ).
Here we  show that this negative   energy is of the order of the Planck mass, 
i.e. smaller than $ O( \l \k) $.
This characteristic  is present   in the RST model as well, where negative energy is carried out by a shock wave (the ``thunderpop") at the endpoint of 
black hole evaporation. As shown below, here the analogue endpoint 
 wave is smeared  in a Planck time.

We start by considering the particular example of the constant incoming flux 
given by   eq. \eecc .
Using the eqs. \ppmm , \energii , \enebb \ and \eennn \
one finds  the following expressions:
\smallskip

\n {\it Subcritical regime} ($a,y \in (0,1)\ $):
\eqn\eeu{
E^{\rm (ii)}_{\rm out}=\l \k \big[-a \ln y -a (1-y)+\ln (1-a+ya)\big]\ ;
}

\n {\it Intermediate regime} ($a  \in (1,\infty )\ , \ \ y\in (y_0, 1) $):
\eqn\eed{
E^{\rm (b)}_{\rm out}=\l \k \big[-a \ln y - (1+b^{-1})^{-1}-\ln (1+b)\big]\ ;
}

\n {\it Supercritical regime} ($a  \in (1,\infty )\ , \ \ y\in (0, y_0) $):
\eqn\eet{
E^{\rm (b)}_{\rm out}=\l \k \big[(1-a) \ln (1-a^{-1})- 
(1+b^{-1})^{-1}- \ln (1 +b^{-1})\big]\ ;
}

\n where
$$
a\equiv {\ep \ov \k} \ ,\ \ \ y\equiv {x^+_0 \ov x^+_1} \ ,\ \ \ 
y_0=1-a^{-1}\ ,\ \ \ b\equiv a(y^{-1}-1)\ .
$$

\n It can be easily seen that the minimum value  of 
 $E^{\rm (ii)}_{\rm out} , \ E^{\rm (b)}_{\rm out}$ given by 
eqs. \eeu --\eet\   is 
$-\l \k $, and it occurs at the point $y=0$ and $a=1$ (corresponding to an incoming flux equal to the critical flux lasting forever).
Thus 
\eqn\lbound{
E^{\rm (b)}_{\rm out}\geq -\l \k\ .
}
This is essentially the same bound  as appears in the RST model. 
 Although we have proved eq. \lbound\ for a constant incoming flux, a similar bound can be obtained in the general case.
  Consider the general expression for
$E^{\rm (b)}_{\rm out}$ in the supercritical
regime (which includes the  case of macroscopic black holes).
It is convenient to write eq. \eennn\ in the form:
\eqn\eenn{ 
E^{\rm (b)}_{\rm out}=
m  -M(x^+_2) + {\l\k p\ov x^-_1} - \l\k  \ln \big(1+{\k\ov \l^2 px_1^+} \big)
-\l \k \ln (x^+_1/x^+_2 ) \ .
 }
From the inequalities
$$
m- M(x^+_2) = \lambda \int_{x^+_2} ^{x^+_1} dx^+ x^+ T_{++} >
\lambda \int_{x^+_2} ^{x^+_1} dx^+ x^+ T_{++}^{\rm cr}=
\l \k \ln (x^+_1/x^+_2 )\ ,
$$
$ {\l\k p\ov x^-_1}> -\l \k $, and 
$$
- \l\k  \ln \big(1+{\k\ov \l^2 px_1^+} \big)
>  -\l\k \ln \big(1+{\k\ov \l^2 px_0^+} \big)
 =   -\l\k \ln \big(1-{ x^-_0\ov  p} \big)> - \l\k  \ln 2\ ,
$$
 we obtain
$$
E^{\rm (b)}_{\rm out}\geq -\l\k -\l\k \ln 2\ .
$$

Next, let us estimate the time interval of the negative energy emission.
For simplicity we will consider the case of  a constant incoming flux.
In Minkowski coordinates $\s^\pm $ the energy momentum tensor
\tttb \ takes the form
\eqn\mintt{
T^{\rm (b)}_{--}= - \k \bigg[
{({\ep\ov\k } -1 )\ov \big( {\ep\ov\k }e^{\l \tau^- }-1\big)^2 }
+{1\ov \big( 1+ {\l^2 px^+_1\ov\k} e^{\l \tau ^-}\big)^2 }\bigg]\ ,\ \ \ 
\l {\tau ^-}\equiv \l \s^- - \ln(\l x^+_1/\k )\ .
}
The shifted Minkowski time $\tau^-$ is such that it starts at zero when
the negative energy emission begins. The second term in eq. \mintt\
is always negligible with respect  to the first one. Since $\ep/\k >1$,
$T^{\rm (b)}_{--}$ is an exponentially decreasing function,
with a damping time interval
of order $\Delta \tau^-= \ha \l^{-1} $, i.e. a ``Planckian"
interval of time (more precisely, $\Delta \tau^-=\ha \l^{-1} (1-{\k \ov\ep})<
\ha \l^{-1}$ ) .

\newsec{ More general distributions of incoming matter}

To   complete the physical picture, let us also give the geometry in region (b) in the case when the
incoming energy-density flux does not stop at $x^+_1$.
Let us assume that $T_{++}(x^+)$ is a smooth function of $x^+$
for all $x^+>x^+_0$, and
define $x^+_1$ as the point at which $T_{++}(x^+)$
becomes less than the critical flux, so that the apparent horizon
becomes time-like after this point.
  Continuity along the line $x_1^-$ requires that
\eqn\sssg{
\Omega^{\rm (b)}(x^+,x^-)=\Omega^{\rm (a)}(x^+,x^-)+F(x^-)\ ,
}
with $F(x^-_1)=0$ and $\Omega ^{\rm (a)}$ as given by eq. \sette .
The expression that generalizes   eqs. \ffff\ and \gggg\ is 
  \eqn\hhhh{
F(x^-)={\k } \ln \bigg( {x^- +\l^{-2}P_+( u) \ov x^- +
\l^{-2}P_+( \hat x^+ ) } \bigg)
+ \l^{-1} M(u)- \l^{-1} M(\hat x^+)\ ,
}
with $u(x^-)$ given as before by the 
branch $x^+_0<x^+<x_1^+$ of the solution $ u=x^+(x^-) $ to the equation
 \eqn\uuuu{
-\l^2 x^+
\big( x^- + \l^{-2} P_+( x^+) \big) = \k  \ ,
}
and $\hat x^+(x^-) $ given by the upper branch  $x^+> x^+_1$.
As in the case of sect. 5, there is no shock-wave discontinuity in
going from region (a) to region (b), since $F'(x^-_1)=0$ (interestingly, 
$\l^{-2} F'(x^-)=   \hat x^+  - u   $, i.e. the distance
between the two points of the apparent horizon corresponding to a given $x^-$).

 The energy-momentum tensor in   region (a) is
as in eq. \ttta\ (since the solution is the same in this region),
and in region (b) one finds   ($p\equiv \l^{-2} P_+(\infty )$)
\eqn\tttbb{
T^{\rm (b )}_{--}(x^- )= -\dd_- ^2\Omega + {\k\ov (x^-+p)^2 }
=\l ^2 {du\ov dx^-} - \l ^2 {d \hat x^+ \ov dx^-} - {\k \ov {x^-}^2 }
+{\k\ov (x^-+p)^2 }\ ,\ \ \ \
}
$$
T^{\rm (c)}_{--}(x^- )=0\ .
$$
This is essentially the   energy-density flux  of eq.  \tttb\  plus an additional (positive energy) contribution of the form \tttii\ 
representing
  reflection of the $T_{++}(x^+) , \ x^+>x^+_1$ on the time-like
apparent horizon. The total mass of the final black-hole geometry will
not vary too much by bombarding it with subcritical energy density. Indeed,
using eqs. \sssg\ and \hhhh\ we find that the final geometry at
$x^+\to \infty $, $x^-\to -p$ is given by
\eqn\mmmbb{
\Omega^{\rm (b)}(x^+,x^-)= -\l ^2 x^+(x^-+p) - \k  \ln \big( -\l ^2 
x^+(x^-+p)\big) +{m_{ f} \ov \l } \ ,
}
$$
m_{ f}=  M(x^+_2) +{\l\k } \ln \big( 1- {P_+(x^+_2)\ov \l ^2 p} \big)\ ,
\ \ \ \ 
x_2^+\equiv u(-p)\ .
$$
This  is  approximately the same static black hole as in the previous
 case, eq. \lmviii , except that now $p$ is slightly different
(since the energy-density flux for $x^+>x^+_1$ is subcritical,
it can be easily seen that $P_+(\infty )-P_+(x^+_1) < k/x^+_1 $).
This difference produces only a tiny (Planck-scale) increase in the final mass $m_f$ with respect to eq. \lmviii .

At first sight, the fact that, for $x^+>x^+_1$, low-energy density
matter reflects on the apparent horizon may seem strange.
However, it must be stressed that this is a quantum effect, since
only a subcritical energy-density flux would reflect.
If, after $x^+_1$, supercritical matter is sent in, the apparent
horizon will become space-like and all but a Planckian bit of energy will
be eaten by the black hole, increasing its size in accordance with the total
energy of the additional  matter.

\newsec{ Outlook and discussion}

Here we have explored the theory  which results  from analytically continuing the
subcritical regime   above the threshold of black hole formation.
In the corresponding semiclassical theory, quantum effects appear in 
various ways, but  the net result is that only 
small alterations    over a   classical picture
appear. Let us summarize the picture:

\smallskip
 \n 1) Collapsing macroscopic matter
(i.e. with incoming energy-momentum tensor
far above   the threshold for black hole formation) forms stable black holes with masses
of the same order as the total imploding energy plus minor emission.
  This involves Hawking radiation at early times and 
a subsequent burst   with   tiny energy (of order of the Planck mass).

\n 2) If the infalling matter has densities not much larger
than the critical one, the situation looks   similar to the conventional
Hawking picture. This is the intermediate regime where a
small black hole is formed and evaporates completely.
 
\n 3) Infalling subcritical matter over an already formed
black hole will be reflected from the apparent horizon with
a small accompanying evaporation.

\n 4) Macroscopic matter falling over a black hole will simply increase
its mass and give rise to a limited emission,    as in 1).

\smallskip

\n The bursts have negative energies of order $\k\l $ and last a short Planckian time $\l ^{-1}$. 
A similar feature appears in the RST model, where the ``Planckian"
negative energy is carried out by a shock wave (the ``thunderpop").
In a sense,  here this shock wave is smoothened-out in a Planckian interval of time.

 The model agrees with the Hawking theory in the region that is not
in causal contact with the apparent horizon (called region (a) in fig. 3). Beyond
this point (region (b)), a  quantum theory of gravity is required in order to predict the outgoing spectrum, since outgoing modes
have Planck frequencies at the moment they arise from the vicinity of the horizon (i.e. about one Planck proper distance from the horizon;
see  refs. \refs{ \thoo , \jacob , \susskin ,  \vercom } ).
Lacking a microscopic theory, some extra phenomenological input is 
needed, and several possibilities have been discussed \refs {\susskin , \ssk,
\hft  , \sussk } .
In the context of this two-dimensional model, this
is naturally realized  in two  scenarios. The first one, described in RST, is based on a quantum field theory with a
boundary at $\Omega= \Omega_{\rm cr }$ (the singularity); the other,
 described here,   
follows from analytic continuation and 
implies a boundary at $\dd_+\Omega=0$ (the apparent horizon).
For the former,    the  physics above the threshold
    reproduces the Hawking model of gravitational collapse,
and thus   leads to information loss. 
As we have seen, this physics  is not  analytically
connected to the subcritical regime: 
    an $S$-matrix constructed on the basis of the subcritical
theory would not describe this conventional approach.\foot{
We stress that we have continued the expressions for the energy-momentum tensor, since a satisfactory $S$-matrix formalism in 1+1 dimensions is, unfortunately, still lacking  (despite some interesting attempts \verli ).}  
 In RST the matching with the vacuum 
is made at the price of a   shock-wave discontinuity. 
In the present model    there are no 
shock-wave discontinuities between the different regions, which are smoothly
matched.

Why is the final geometry stable? 
The vanishing of the energy-momentum tensor at infinity requires 
--just as when the Boulware vacuum is adopted \boulw -- a substantial modification
of the geometry near the line $x^-=-p$. As we have seen in sect. 5.2,
this is exactly what is happening. In the allowed space-time region,
 $-\l ^2 x^+(x^-+p)>\k$, the geometry
is essentially the same as the classical black hole geometry.
Only at $ -\l ^2 x^+(x^-+p)\ll e^{-m/ \l \k} $ is the geometry   significantly
 modified, but this lies outside the boundary.
The  difference with
  the conventional results stems  indeed from the small trapped region (of Planckian
proper length) in the causal past of  null infinity. This eliminates the instability of the hole, inhibiting radiation --at transplanckian frequencies-- that would otherwise cause its evaporation (experiencing in the process a tremendous red shift).

There is a close relation   between the boundary at the apparent horizon
--dictated by analytic continuation-- and the stretched horizon proposed
by Susskind et al. \susskin\ as an effective boundary to external observers,  whose
dynamics could determine the modification to the Hawking radiation 
(the apparent horizon always lies inside the stretched horizon and it coincides with it after the incoming flux   terminates \hori\  --in the present model the stretched horizon is just given by 
$-\l^2 x^+\big( x^- + \l^{-2} P_+(\infty )\big)=\k $~).
 This dynamics could be quite different in the 3+1 dimensional physics,
 where it is possible to have classical scattering   without
black hole formation (e.g. in terms of the impact parameter). 
In 1+1 dimensions there is no classical scattering without black holes;  
the threshold is a pure quantum effect (a simple model in 3+1 dimensions is investigated in \black ).
The unconventional  results of this two-dimensional model  
 provide, however,  a simple example of how an analytic continuation
from a subcritical regime --natural in   an $S$-matrix approach--
may shed light on  black-hole behavior in a theory that
does not lose quantum coherence.


\bigskip
\bigskip


\bigskip\bigskip

   \listrefs
\vfill\eject
 \end